\documentstyle[aps,preprint]{revtex}
\begin{document}
\draft
\title{Statistical Properties of Random Banded Matrices with
Strongly Fluctuating Diagonal Elements}
\author{Yan V. Fyodorov$^{1,*}$ and Alexander D. Mirlin$^{2,*}$ }
\address {$^1$ Fachbereich Physik,
Universit\"{a}t-Gesamthochschule Essen,
 Essen 45117, Germany}
\address{$^2$
 Institut f\"{u}r Theorie der Kondensierten Materie,
  Universit\"{a}t Karlsruhe, 76128 Karlsruhe, Germany}
\date{\today}
\maketitle
\tighten

\begin{abstract}
The random banded matrices (RBM) whose diagonal elements fluctuate
much stronger
than the off-diagonal ones were introduced recently by Shepelyansky
as a convenient model for coherent propagation of two interacting
particles in a random potential. We treat the problem analytically
by using the mapping onto the same supersymmetric
nonlinear $\sigma-$model that
appeared earlier in consideration of the standard RBM ensemble, but
with renormalized parameters. A Lorentzian form of the local density
of states and  a two-scale spatial structure of the eigenfunctions
revealed recently by Jacquod and Shepelyansky are confirmed by
direct calculation of the distribution of eigenfunction components.
\end{abstract}

The ensemble of Random Banded Matrices (RBM) can be  generally
described as that of large $N\times N$ matrices having nonzero
elements effectively within some wide band of the width $b\gg 1$
around the main diagonal.
Such a structure naturally appears in various physical
contexts, and serve as a useful model in Quantum Chaos\cite{Izr},
atomic physics \cite{Flam}
and solid state physics \cite{FMrev}. Due to this fact a lot of
efforts were spent in order to study different kinds of RBM, both
numerically \cite{num1,num2} and analytically \cite{an1,an2}. In
particular, it was found that  the problem can be
mapped onto a supersymmetric  $1d$
nonlinear $\sigma$--model introduced in \cite{EF}, provided all matrix
elements within the band are independent and distributed around zero.
More precisely, the mapping was shown to exist for those matrices
whose variance $\langle |H_{ij}|^2 \rangle$ was dependent on the
distance $|i-j|$ from the main diagonal: $\langle |H_{ij}|^2
\rangle=b^{-1}f(|i-j|/b)$ where the function $f(r)$ is of the order
of unity when $r\lesssim1$ and decreases exponentially (or faster) at
$r\gg  1$.

Quite recently, Shepelyansky \cite{Dima} argued that a very
interesting problem of two interacting particles propagating in a
quenched random potential can be effectively mapped onto a class of
RBM whose diagonal elements $H_{ii}$ fluctuate much stronger
than off-diagonal ones: $\langle |H_{ii}|^2 \rangle/\langle
|H_{ij}|^2 \rangle
\propto b\gg1$. Using this kind of mapping, Shepelyansky predicted a
considerable interaction-assistant enhancement of the two-particle
localization length as compared with the localization length of one
particle in the same random potential. This conclusion was confirmed
later on by Imry \cite{Imry}
who employed the Thouless scaling block picture bypassing the
mapping to RBM.
Subsequent numerical studies \cite{Pr} also confirmed main
qualitative result by Shepelyansky, but revealed some deviations from
the predicted behavior of the two-particle localization length,
which were
attributed to oversimplified statistical assumptions concerning RBM
elements
in Shepelyansky construction. Nevertheless, it is clear that
Shepelyansky RBM model (SRBM) catches adequately at least some of the
important features of the original physical problem and thus
deserves more detailed study.

In a very recent paper \cite{JS}, Jacquod and Shepelyansky
presented their detailed numerical results on statistical properties
of SRBM. They revealed a peculiar structure of eigenfunctions
$\Psi_{\alpha}$, consisting of a set of large spikes separated by
regions of relatively small amplitude. Such a ``sparse'' spatial
arrangement shows up in a difference  between the localization
length
$l$ related to the rate of a spatial decay of an  eigenfunction
envelope, $l=\frac{1}{n}\lim_{n\to
\infty}\ln{|\Psi_{\alpha}(0)\Psi_{\alpha}(n)|}$,
and the length $\xi$ defined as the  participation ratio,
$\xi=\left(\sum_n |\Psi_{\alpha}(n)|^4\right) ^{-1}$. For a
conventional ``dense''
eigenfunctions these two lengths are expected to be of the same
order of magnitude, whereas for SRBM it was found that $l\gg \xi$.
Another interesting feature making  SRBM different from earlier
studied cases is that in any given realisation of the disorder the
local density of states (LDOS) defined as
\begin{equation}\label{dos}
\rho(E,n)=\sum_{\alpha}\mid \Psi_{\alpha}(n)\mid ^{2}\delta
(E-E_{\alpha})
\end{equation}
was found to follow the simple Lorentzian form with a width
$\Gamma\propto W_b^{-1}$ independent of the parameter $b$, where
$W_b\gg 1$ determines the scale of fluctuations of the
diagonal elements $H_{nn}$.
To this end it is appropriate to mention that the Lorentzian form of LDOS
was earlier found to be typical for RBM with linearly increasing
mean value
of the diagonal elements: $\langle H_{nn}\rangle=\beta n$
\cite{Wig,Flam}.

In the present article we show that Shepelyansky RBM model can be
again mapped onto the standard one-dimensional nonlinear
$\sigma-$model with modified parameters. This fact allows us to
reproduce analytically most of peculiar features of the SRBM
discussed above.

We consider the random Hermitian\cite{note}
matrix $H_{ij}=W_i \delta_{ij}+H^{(0)}_{ij}$ , where the matrix
$H^{(0)}_{ij}$ is a standard RBM characterized via the variances:
$J_{ij}=\langle H^{(0)*}_{ij}H^{(0)}_{ij}\rangle=
\frac{1}{b}f(\mid i-j\mid/b)$, normalized in such a way that
$\sum_{r=-\infty}^{\infty}\frac{1}{b}f(r/b)=1$.
This normalization ensures that the width of the energy spectrum of
the matrix $H^{(0)}_{ij}$ is of order unity in the limit
$b\to\infty$.
The parameters $W_i$
are assumed to be independently distributed around zero according
to the probability density
 ${\cal P}(W)=\frac{1}{W_b}h(W/W_b)$ where $h(\tau\sim 1)\sim 1$
and $\int_{-\infty}^{\infty} h(\tau) d\tau =1$.

Depending on the value of $W_b$, the following three regimes should be
distinguished:
\begin{itemize}
\item[i)] $W_b\ll 1$. The ensemble is completely equivalent to the
conventional RBM ensemble; diagonal matrix elements do not play an
essential role;
\item[ii)] $W_b\gg\sqrt{b}$. Perturbative regime. The eigenststes can
be approximated by the eigenstates of the diagonal matrix
$W_i\delta_{ij}$, which are localized on single sites. The
non-diagonal term $H_{ij}^{(0)}$ in the Hamiltonian can be then
treated via the perturbation theory;
\item[iii)] $1\ll W_b\ll \sqrt{b}$. Intermediate regime.
It is just the regime shown to be relevant for the problem of two
interacting particles in random potential \cite{Dima}.This case is
our main concern in the present article.
\end{itemize}

We are going to characterize eigenfunction statistics via the
following correlation function, see \cite{FMrev,an2}:
\begin{eqnarray}
{\cal K}_{l,m}&=&\langle n|(E+i\eta-\hat{H})^{-1}|n\rangle^l
\langle n|(E-i\eta-\hat{H})^{-1}|n\rangle^m  \nonumber\\
&=&\frac{i^{m-l}}{l!m!}\int\prod_i d\Phi_i \:
(S_{n,1}^*S_{n,1})^l(S_{n,2}^*S_{n,2})^m \nonumber\\
&\times &
\exp\{i\sum_i\Phi_i^\dagger[(E-W_i+i\eta\Lambda]L
\Phi_i-i\sum_{\langle ij\rangle}H_{ij}\Phi_i^\dagger L
\Phi_j\}
\label{corr}
\end{eqnarray}
where $\Phi_i^{\dagger}=(S^{*}_{i,1},\chi^{*}_{i,1},S^*_{i,2}
,\chi^*_{i,2})$, with $S_{i,p}$ and $\chi_{i,p}$ being complex
commuting and Grassmannian variables, respectively.
The $4\times 4$ matrices $\hat{\Lambda},\hat{L}$
are diagonal and have the following structure:
$ \hat{\Lambda}=\mbox{diag}(1,1,-1,-1);\quad
\hat{L}=\mbox{diag}(1,1,-1,1)$.

Let us first calculate such a correlation function for arbitrary
{\it fixed} value
of the potential $W_n$ in the observation point $n$,
performing both averaging over $H^{(0)}_{ij}$
and over all  $W_j$ with $j\ne n$ (the latter averaging we denote as
$\langle ... \rangle_W$ henceforth). Repeating all the
necessary steps
outlined in \cite{an2} and presented in more details in
\cite{FMrev} one expresses the correlation function in terms of the
integral over the set of supermatrices $R_i=T^{-1}_i P_i T_i$, where
the supermatrices $P_i$ are $4\times 4$ block-diagonal ones and
$T_i$ belong to the graded coset space
$U(1,1/2)/U(1/1)\times U(1/1)$. The resulting expression looks as
follows:
\begin{eqnarray}
{\cal K}_{l,m}(E,n;\eta)&=&\frac{i^{l-m}}{l!m!}\int\prod dR_i{\cal
F}(R_n)
\exp{[-i\eta\sum_{i}\mbox{Str} R_i\Lambda-{\cal L}(R)]}\\ \nonumber
{\cal L}(R)&=&\frac{1}{2}\sum_{ij}(J^{-1})_{ij}Str R_iR_j-
\sum_{i=1}^{N}\ln\langle \mbox{Sdet}^{-1}(E-W-R_i)\rangle_W\\ \nonumber
{\cal F}(R_n)&=&\sum_{k=0}
\left(\begin{array}{c}l\\k\end{array}\right)
\left(\begin{array}{c}m\\k\end{array}\right)
G_{n,11}^{l-k}G_{n,33}^{m-k}G_{n,13}^{k}G_{n,31}^{k}
\frac{\mbox{Sdet}G_n}{\langle \mbox{Sdet}(E-W-R_n)^{-1}\rangle}_W
\end{eqnarray}
where $G_n=(E-W_n-R_n)^{-1}$. The notations $\mbox{Sdet}$ and
$\mbox{Str}$
stand for the graded determinant and graded trace, correspondingly.

The integral over the matrices $P_i$ can be calculated in the limit
$b\gg 1$ by the saddle-point method. The saddle-point solution
$P_i\equiv P_s$ is diagonal and independent of the index $i$. The
diagonal matrix elements
$d$ satisfy the following equation:
\begin{equation}\label{d}
d=\int \frac{dW}{W_b}h(W/W_b)\frac{1}{E-W-d}
\end{equation}
Equations of similar type  appeared in earlier studies of full
random matrices with preferential diagonal\cite{Pas}, most recently
in \cite{Zel,Prus}, and known as Pastur equation.
The subsequent analysis depends on the value of the parameter $W_b$
characterizing the strength of the diagonal disorder. If $W_b\ll 1$,
we can neglect $W$ in denominator in the r.h.s. of eq.(\ref{d}). Then
the diagonal matrix elements distribution $h(\tau)$ drops out from the
formulae, and the results are precisely the same as for the
conventional RBM ensemble. In the present
paper we are interested in the opposite case, $W_b\gg 1$. Then one
obviously has $|d|\propto W_b^{-1}\ll W_b$ and to the leading order
in $W_b^{-1}$ one finds:
\begin{equation}
 d=\frac{1}{W_b}{\cal P}\int \frac{dW}{E-W}h(W/W_b)\pm
i\frac{\pi}{W_b}h(E/W_b)\equiv\mbox{Re}\,d\pm i\mbox{Im}\,d\ ,
\label{101}
\end{equation}
where ${\cal P}$ stands for the principal value of the integral.

As usual, the correct saddle-point solution is equal to
$P_s=(\mbox{Re}\,d)\,\hat{I}+i (\mbox{Im}\,d)\,\Lambda $. In order
to find
the region of applicability of the saddle-point method we expand the
functional ${\cal L}(R)$ around the saddle-point value and
calculate the corrections due to gaussian fluctuations.
The latter turn out to be of the order of $\overline{(\delta
P_i)^2}\propto b^{-1}$. Comparing this value with the saddle point
one $P_s^2\sim d^2
\sim W_b^{-2}$ we conclude that corrections are small as long as
$W_b^2\ll b$.
Thus, our calculation is completely legitimate everywhere in the
{\it nonperturbative} regime $1\ll W_b\ll b^{1/2}$ which is just
the case relevant for the physical applications of SRBM \cite{Dima}.

Introducing the set of matrices $\hat{Q}_i=-iT_i^{-1}\Lambda T_i$
and using the identity
$$
(E-W_n-R_n)^{-1}\mid_{P=P_s}=\frac{(E-W-\mbox{Re}\,d)\,\hat{I}-
(\mbox{Im}\,d) \,\hat{Q}}
{(E-W-\mbox{Re}\,d)^2+(\mbox{Im}\,d)^2}$$
one arrives at the following expression for the correlator (\ref{corr}):
\begin{equation}\label{sigma1}
{\cal K}_{l,m}(E,n;\eta)=\int d\mu(Q){\cal F}_{l,m}(Q)e^{-S(Q)}
\end{equation}
where the action
\begin{equation}\label{sigma2}
S(Q)=-\frac{\gamma}{2}Str\sum_{i=1}^{N-1}Q_iQ_{i+1}+i\epsilon\sum_{i=1}^{N}
Str(Q_i\Lambda)
\end{equation}
defines the standard one-dimensional nonlinear graded
$\sigma-$model on a lattice characterized by the coupling constant
$\gamma=\left(\Gamma/2\right)^2\sum_r J(r)r^2$ and the effective level
broadening $\epsilon=(\Gamma/2)\eta$, the parameter $\Gamma/2$
being equal to $\Gamma/2=\mbox{Im}\,d\propto W_b^{-1}$.

The nonlinear $\sigma-$model defined in
eqs.(\ref{sigma1},\ref{sigma2}) was studied in much details in
\cite{FMrev,an1,an2}. In particular, one can immediately extract the
value of the localization length $l$ which is known to be
proportional to the coupling constant: $l=4\gamma\propto
b^2/W_b^2$,
in full agreement with the results of \cite{Dima}. Another
quantity that can be most easily calculated is the mean local DOS
defined in eq.(\ref{dos})
and given by:
\begin{eqnarray}
\rho(E,n)&=&\frac{1}{\pi}\mbox{Im}\langle
n|(E-i\eta-\hat{H})^{-1}|n\rangle\mid_{\eta\to 0}\nonumber\\
&\equiv&
\frac{1}{\pi}\mbox{Im}\,{\cal K}_{l=0,m=1}(n,E;\eta\to 0)\nonumber\\
&=&\displaystyle{\frac{1}{\pi}\frac{\mbox{Im}\,d }
{(E-W_n-\mbox{Re}\,d)^2+(\mbox{Im}\,d)^2}}
\label{meandos}
\end{eqnarray}

We conclude therefore that typically the local DOS is a  Lorentzian
centered around $E=W_n+ \mbox{Re}\,d$ with the width
\begin{equation}\label{width}
\Gamma/2=\mbox{Im}\,d=\frac{\pi}{W_b}h\left({E\over W_b}\right)
\propto \frac{1}{W_b}
\end{equation}
as it was indeed found in the numerical studies \cite{JS}. The center
of this Lorentzian is shifted from the local value of the random
potential, $W_n$, by the amount $\mbox{Re}\,d\sim W_b^{-1}$. The shift
is small compared to the typical values of $W_n\sim W_b$.

Knowing all the correlators ${\cal K}_{lm}(E,n;\eta)$, one
can extract the full set of the eigenfunction moments
$P_q(E)=\overline{\sum_n|\Psi_{\alpha}(n)|^{2q}}$
and, finally, the whole probability distribution of the
eigenfunction amplitude
$|\Psi_{\alpha}(n)|^2$ \cite{an2}. Straightforwardly repeating all
the necessary
steps one finds that all moments $P_q(E)$ for SRBM are
proportional to the corresponding
moments for standard RBM at the same values of the  parameters
$N$ and $\gamma$. Putting the energy $E$
to be zero for the sake of simplicity, one obtains:
\begin{equation}\label{moments}
P_q(E=0)\mid_{\mbox{SRBM}}=\left\langle
\frac{1}{[W^2+(\Gamma/2)^2]^q}
\right\rangle_{W}P_q(E=0)\mid_{\mbox{standard RBM}}
\end{equation}
It is well known that for the standard RBM in the localized regime
$N\gg \gamma$ one has $P_q\propto \gamma^{1-q}$ \cite{an2}. The
relation eq.(\ref{moments})  tells us that
$P_q\mid_{\mbox{SRBM}}
\propto (\Gamma^2 \gamma)^{1-q}\propto (W_b^2/l)^{q-1}$.
In particular, for the participation ratio
$\xi=P_2^{-1}$ one has
$\xi\propto l/W^2\propto b^2/W^4$, thus proving the abovementioned
difference between $\xi$ and $l$ discovered by Jacquod and
Shepelyansky \cite{JS}.

The relation between the moments eq.(\ref{moments}) allows one to express
the distribution function of the normalized eigenfunction amplitude
$y=N|\Psi^2|$ for the Shepelyansky RBM model in terms of that for
the standard RBM:
\begin{equation}\label{dis}
{\cal P}^{SRBM}(y)=\frac{\Gamma^3}{4W_b}\int_{1}^{\infty} \frac{du
u}{(u-1)^{1/2}}
h\left(\frac{\Gamma}{2W_b}\sqrt{u-1}\right){\cal
P}^{RBM}\left(y\frac{\Gamma^2}{4}u\right)
\end{equation}

The actual form of the function ${\cal P}^{RBM}(y)$ depends on the
scaling ratio $N/\gamma$ and can be found in Refs.\cite{FMrev,an2}
It takes a simple form in both localized limit $N\gg
\gamma$ and delocalized limit $N\ll \gamma$. For example, for the
latter case
${\cal P}^{RBM}(y)=e^{-y}$ and therefore one gets:
\begin{equation}\label{del}
{\cal P}^{SRBM}(y)=-W_b\frac{\partial}{\partial
y}\left[\frac{e^{-y\Gamma^2/4}}{\sqrt{y}}\int_{-\infty}^{\infty}
dze^{-z^2}h\left(\frac{z}{\sqrt{y}W_b}\right)\right]
\end{equation}

This distribution clearly displays the presence of two scales. All
moments $\langle y^q\rangle\equiv P_q$
with $q\ge1$ are dominated by the region
where $y \Gamma^2\sim 1$ (correspondingly, $|\Psi^2|\sim W_b^2/N$),
whereas
the normalization integral  $\int {\cal P}(y) dy$ is dominated by
the values $yW_b^2\sim 1$, where eigenfunction amplitude is small:
$|\Psi^2|\sim (W_b^2 N)^{-1}$. This result corresponds the following
picture of a typical
delocalized eigenstate:  the eigenfunction consists of isolated
peaks with typical amplitude $|\Psi^2|\sim W_b^2/N$ separated by
regions of a typical
spatial extent $L\sim W_b^2$ filled in with a low amplitude
components $|\Psi^2|\sim (W_b^2N)^{-1}$.

Essentially the same picture holds for the regime of strong localization
$l\ll N$. Here any eigenstate has a profile which is exponentially small
outside the spatial region of the size $l$. However, within this
region there are isolated spikes of the amplitude $|\Psi^2|\sim W_b^2/l$
separated by the low amplitude regions with a typical extent $L\sim
W_b^2$ where the wave function amplitude is small:
$|\Psi^2|\sim (W_b^2 l)^{-1}\sim 1/b^2$.

One can also calculate for SRBM other quantities known for
standard RBM.
For example, one can be interested in level-to-level fluctuation of
the  participation ratio $\xi$.
Performing such a calculation one finds that if one normalizes
the inverse participation ratio by its mean value $\langle P_2\rangle
\propto W^4/b^2$,
then the distribution of the quantity $z=P/\langle P_2\rangle$
coincides exactly with the distribution found for standard RBM in
\cite{an1}. This fact suggests that  envelopes
 of high amplitude peaks in SRBM are typically quite similar to
envelopes
of eigenfunctions in conventional RBM, after appropriate rescaling.

Our last comment concerns the spectral correlator $Y_2(\omega)=\langle
\rho(E)\rho(E+\omega)\rangle$ for SRBM. It is easy to satisfy
oneself that
everywhere in the region $W_b\ll b^{1/2}$
the function $Y_2(\omega)$ for SRBM coincides with that known for
RBM, as long as two-level separation $\omega$ is small in comparison
with the spectral width $\Gamma$ of the LDOS: $\omega\ll W_b^{-1}$.
Recently Prus and Sivan\cite{Prus}  addressed the same question
for the case of full matrices $b=N$. They found that the
functions $Y_2(\omega)$
is given by the same expression as that for the standard Gaussian
matrices everywhere
in the region $\omega\ll 1/W_b$ as long as $W_b\ll N^{1/2}$. The
latter result is also in correspondence with numerical studies by
Lenz et.al \cite{Lenz}
who found that the crossover from the Wigner-Dyson statistics to
the Poissonian one occurs at the scale $W_b^2\sim N$.
This scale is much larger than the scale $W_b\sim 1$ necessary to
induce changes in the form of the mean density\cite{Lenz,Shap}.
Thus, the sparse structure of the eigenstates  discussed above has
no effect on the spectral statistics at relatively low frequency
$\omega$, as long as the system stays
well in the {\it nonperturbative} regime.

Y.V.F  is grateful to D.Shepelyansky for attracting his interest to
the problem and for the useful communication and to O.Prus for
informing him on his unpublished results \cite{Prus}. This work was
suported by SFB 237 (Y.V.F.) and SFB 195 (A.D.M.) der Deutschen
Forschungsgemeinschaft.


\begin{references}
\bibitem{note} For the sake of convenience we present our results
for Hermitian matrices. All results can be straightforwardly
generalized to symmetric case as well.
\end{references}
\end{document}